\documentclass{article}

\usepackage{arxiv}

\usepackage[utf8]{inputenc} % allow utf-8 input
\usepackage[T1]{fontenc}    % use 8-bit T1 fonts
\usepackage{hyperref}       % hyperlinks
\usepackage{url}            % simple URL typesetting
\usepackage{booktabs}       % professional-quality tables
\usepackage{amsfonts}       % blackboard math symbols
\usepackage{nicefrac}       % compact symbols for 1/2, etc.
\usepackage{microtype}      % microtypography
\usepackage{xcolor}         % colors
\usepackage{graphicx}
\usepackage{svg}
\usepackage{amsmath}
\usepackage{multirow}

\title{ProtFAD: Introducing function-aware domains as implicit modality towards protein function prediction}

\author{
\textbf{Mingqing Wang}\textsuperscript{1,3}, \textbf{Zhiwei Nie}\textsuperscript{2,3}, \textbf{Yonghong He}\textsuperscript{1}, Athanasios V. Vasilakos\textsuperscript{4}, \textbf{Zhixiang Ren}\textsuperscript{3}\thanks{Corresponding author} \\
\textsuperscript{1}Tsinghua University, China \quad \textsuperscript{2}Peking University, China \quad \\
\textsuperscript{3}Peng Cheng Laboratory, China \quad
\textsuperscript{4}CAIR, University of Agder, Norway\\
\texttt{renzhx@pcl.ac.cn} \\
}

\begin{document}

\maketitle

\begin{abstract}
Protein function prediction is currently achieved by encoding its sequence or structure, where the sequence-to-function transcendence and high-quality structural data scarcity lead to obvious performance bottlenecks.
Protein domains are "building blocks" of proteins that are functionally independent, and their combinations determine the diverse biological functions.
However, most existing studies have yet to thoroughly explore the intricate functional information contained in the protein domains.
To fill this gap, we propose a synergistic integration approach for a function-aware domain representation, and a domain-joint contrastive learning strategy to distinguish different protein functions while aligning the modalities.
Specifically, we align the domain semantics with GO terms and text description to pre-train domain embeddings.
Furthermore, we partition proteins into multiple sub-views based on continuous joint domains for contrastive training under the supervision of a novel triplet InfoNCE loss. 
Our approach significantly and comprehensively outperforms the state-of-the-art methods on various benchmarks, and clearly differentiates proteins carrying distinct functions compared to the competitor. 
Our implementation is available at https://github.com/grizzledtroubadour/ProtFAD.
\end{abstract}

%Protein function prediction; Protein domain; Deep learning; Function Priors; Contrastive Learning
% ------------------------------------------------------------------------------------------
\section{Introduction}
\label{sec:intro}

% figure 1: overall motivation
\begin{figure}
\centering
\includegraphics[width=1\textwidth]{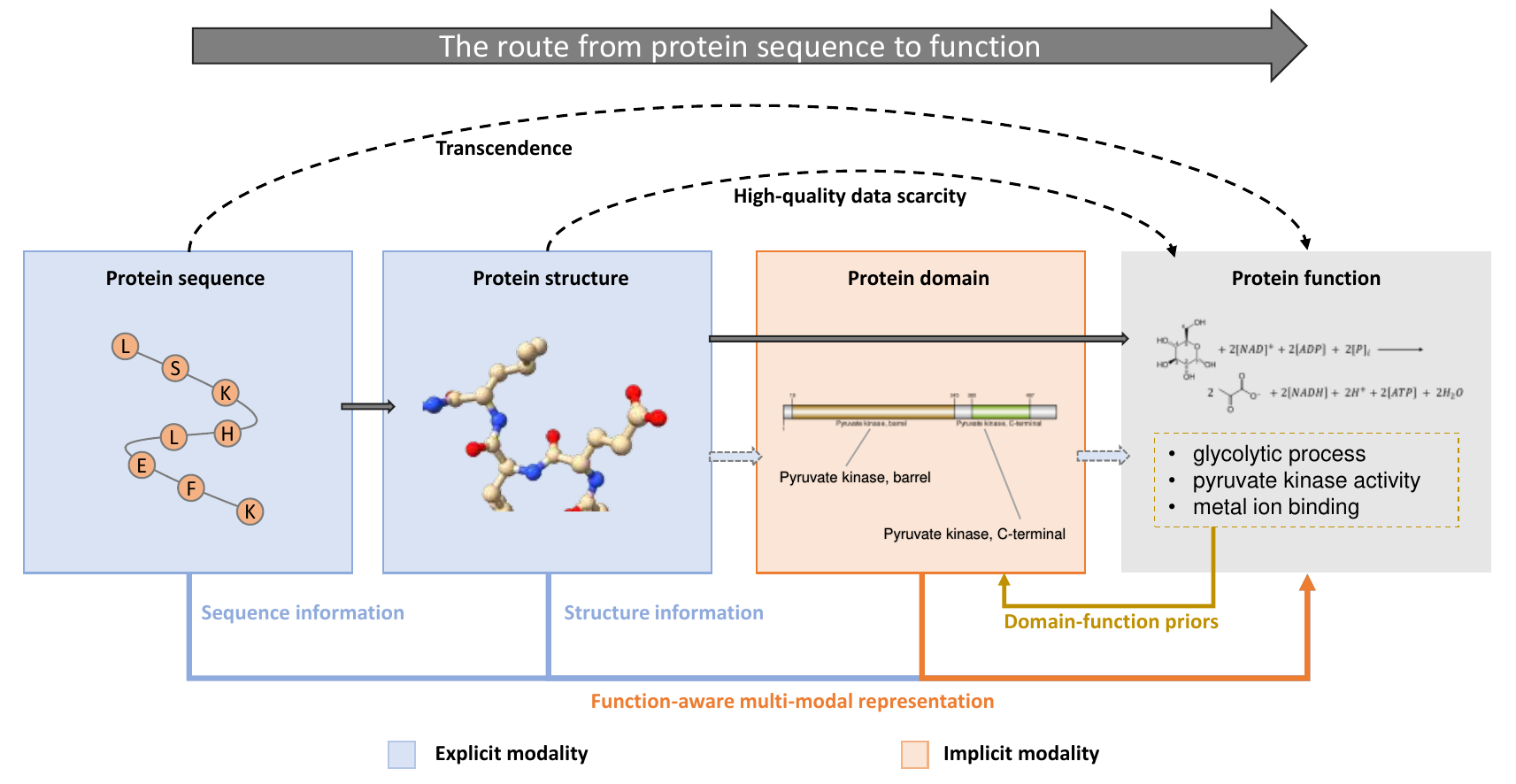}
\caption{
The motivation and methodology of our method.
The performance bottlenecks of existing protein function prediction methods are rooted in sequence-to-function transcendence and high-quality structural data scarcity.
On the path from protein sequence to function, unlike explicit modalities such as sequence or structure, the protein domain is a function-oriented implicit modality carrying domain-function priors.
By introducing this implicit transition modality, the sequence or structure modality and protein function can be bridged in a tangible way.
}
\label{fig:motivation}
\end{figure}

% 首先说明蛋白质的重要性，介绍深度学习工作在蛋白质方面的重要突破；
% 但是人类对蛋白质的理解还有着鸿沟，而蛋白质表征学习/性质预测是跨越这一鸿沟的关键。
Proteins play a pivotal role in the biological processes of living organisms, contributing to cell structure, functionality, signal transduction, and enzymatic reactions \cite{karplus2005molecular,benkovic2003perspective,pawson2000protein}. 
%Understanding proteins is crucial for various fields, including biochemistry, molecular biology, and medicine. 
With the development of deep neural networks, remarkable breakthroughs have been achieved in the research of proteins, including in protein-ligand binding \cite{diffdock}, variant effect prediction \cite{alphamissense, esm1v}, \textit{de~novo} protein design \cite{rfdiffusion}, etc. 
Despite these advancements, there remains a substantial gap in our understanding of proteins, particularly in deciphering the intricate relationships between protein sequence, structure, and function. Protein function prediction has emerged as a focal point in addressing this gap, aiming to identify the specific roles and activities of proteins within biological systems \cite{survey1, survey2}. 
%This research has the potential to revolutionize our understanding of proteins and pave the way for discoveries in biology and medicine.

% 现有的蛋白质表征学习/性质预测工作主要使用蛋白质序列、结构数据；
% 分别解释使用蛋白质序列或结构数据对蛋白质进行表征的不足。

As shown in Figure \ref{fig:motivation}, current computational methods often rely on using sequence or structure data for protein function prediction \cite{cdconv, esm-gearnet, lm-gvp}.
Protein sequences, which are linear chains of amino acids, contain valuable information about the composition and order of amino acids in a protein. However, the transcendence from sequences to protein functions leads to a challenge to the generalizability and interpretability of sequence-based methods for protein function prediction.
On the other hand, protein structures provide precise spatial information, which fundamentally determines the protein functions. However, the scarcity of high-quality structural data leads to performance bottlenecks in such methods \cite{ma2022enhancing, heal}.
Even though protein structure prediction methods \cite{alphafold2, esm2struct,abramson2024accurate} have made breakthrough progress, their inherent errors and insensitivity to structural rearrangements caused by mutations \cite{yin2022benchmarking,stevens2022benchmarking,mcdonald2023benchmarking} limit the easy acquisition of high-quality structural data.

%These biases may limit the comprehensive understanding of protein function, and prompt a critical reevaluation of existing approaches.

% 介绍蛋白质域对于蛋白质表征的重要性；
% 介绍现有的基于蛋白质域的蛋白质表征方法，并说明现有方法的不足；

Protein domains are distinct structural and functional units within a protein that can exist and function independently, as shown in Figure \ref{fig:motivation}. 
Given that these domains usually determine the specific function of the protein, such as binding to other molecules or catalyzing chemical reactions, we can treat it as a function-oriented implicit modality carrying domain-function priors.
Unlike explicit modalities such as sequence or structure, the protein domain is an implicit transition modality between structure and function.
In recent years, studies have shown that protein domains provide valuable information for protein representations, enabling better predictions of protein function and behavior \cite{netgo2, domainpfp}. 
Existing methods \cite{sdn2go, graph2go, torres2021protein, netgo, golabeler} employ protein domains as a source of complementary functional information in ensemble frameworks. However, these methods only utilize the coarse-grained information of protein domains, such as their type or quantity, while overlooking the functional information contained within protein domains. 
In addition, some studies \cite{domainpfp, dom2vec} learn vector representations of protein domains with higher information density. However, these approaches only focus on mining information from protein domains, neglecting to learn a robust multi-modal representation, and thus missing out on a substantial amount of important protein information.

% 最后提出本文的方法，即使用包含功能信息的域嵌入来增强现有的基于序列和结构的蛋白质表征
% 首先要从已有的蛋白质的域和GO terms中挖掘域所包含的功能信息，并训练出功能感知的域嵌入,（然后还使用了一个域box位置编码和一个域注意力来处理域位置信息和交互作用）

% 感觉这里可以不用写的这么平凡，可以详细说明实验部分的分析，并且强调性能的提升主要来源于domain/function信息的融入；再可以引入实验过程中的噪声现象（有条件可以利用可视化证明该噪声现象真实存在），再提到本文解决该噪声现象的方法，最好再利用可视化去证明joint domain之间不存在噪声。

To overcome the limitations mentioned above, we propose augmenting existing sequence-based and structure-based protein representations with domain embeddings containing function priors, as shown in Figure~\ref{fig:overall}. 
Specifically, we associate domains and Gene Ontology (GO) terms from various proteins and further calculate the domain-function probability from the constructed data pairs. Furthermore, we collect the text description of domains.
By training domain vocabularies with constructed pseudo labels and the semantically consistent loss, we incorporate the function priors into the domain embeddings (i.e. function-aware domain embeddings, FAD). 

\begin{figure}
\centering
\includegraphics[width=1\textwidth]{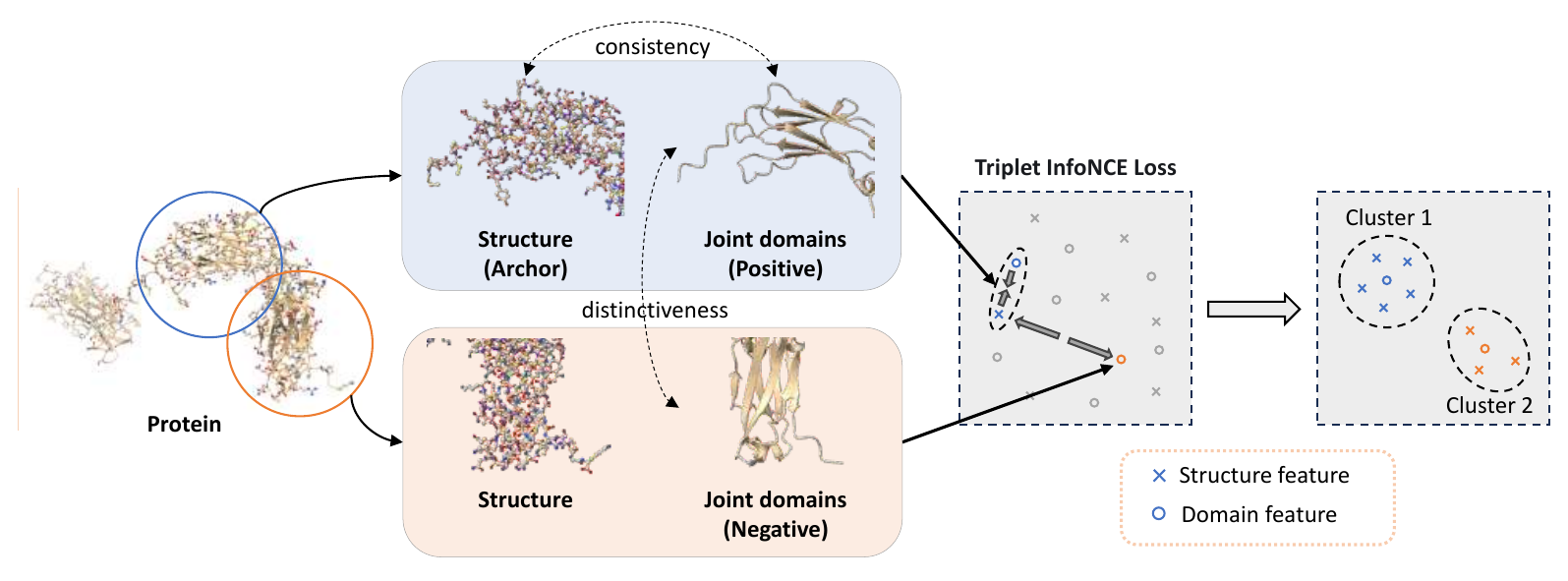}
\caption{
\textbf{Protein domain-joint contrastive learning.}
Here, we demonstrate domain-joint contrastive learning between structure and joint domains, which distinguishes different protein functions while aligning the modalities.
Contrastive learning between sequence and joint domains is similar.
}
\label{fig:domain-joint}
\end{figure}

% 然后使用提出的protein domain-joint contrastive learning来融合蛋白质序列、结构和域信息
% 这里要强调同源蛋白序列、结构相似度高，给对比学习引入噪声的问题，再说明提出的domain-joint contrastive learning。
Subsequently, we build a multi-modal function-prediction framework that integrates sequences, structures, and function-aware domains, including a domain attention module that extracts the functional representations of joint domains. To align the modalities while distinguish different functions and overcome the inevitable noise in protein data (e.g. caused by homologous proteins with similar functions), we propose the domain-joint contrastive learning strategy with a novel triplet InfoNCE loss, as shown in Figure~\ref{fig:domain-joint}. 
This multifaceted representation, bridging sequence or structure with protein functionality, holds promise for enhancing the robustness and interpretability of protein function prediction.

The main contributions of this paper are as follows:

\begin{itemize}
\item We introduce a protein function-oriented implicit modality, i.e., function-aware domain embeddings, to bridge the gap between sequence or structure modality and protein functions.
\item We propose a novel domain-joint contrastive learning strategy that enables neural networks to align different modalities while distinguishing different protein functions.
\item Our approach significantly outperforms state-of-the-art competitors on multiple benchmarks across the board, additionally perceiving proteins carrying specific functions.
\end{itemize}

% ------------------------------------------------------------------------------------------
\section{Related work}
\label{sec:rel}

In this section, we review previous studies about sequence-based, structure-based, and multi-modal protein representation learning respectively.
Related studies of the protein domain are also summarized and introduced.
We discuss the strengths and limitations of each approach and highlight directions for our research below.

\textbf{Protein representation learning.}
Protein representation learning plays a crucial role in protein research, such as protein function prediction \cite{deepfri, heal}, protein-protein interaction prediction \cite{kang2023bbln}, and drug discovery \cite{pan2023submdta, wu2024attentionmgt, zhang2023multimodal}. Several approaches have been developed to learn protein representations, leveraging different aspects of protein information. Inspired by natural language processing techniques, protein language models (PLMs) \cite{esm1b, esm2, prottrans} learn to generate meaningful embeddings that encapsulate the hierarchical structure and evolutionary relationships of proteins by training on large-scale protein sequence datasets. \cite{saprot, rao2021msa} integrate additional information (e.g. "structure-aware vocabulary" or family information) to improve the performance of PLMs. 
The structure of a protein directly determines its function. Therefore, more and more approaches focus on training protein representations using structural data. \cite{gvp, ieconv, pronet, cdconv} introduces novel network operators to perform both geometric and relational reasoning on efficient representations of macromolecules. \cite{steps, hermosilla2022contrastive, zhang2024pre, gearnet} employ self-supervised learning methods to effectively capture structural information of proteins and learn meaningful protein representations. However, the limited number of available protein structures severely limits the performance of these methods. AlphaFold2 and its open-source data set of predicted protein structures are expected to benefit the model performance by increasing the number of training samples \cite{ma2022enhancing, heal}. However, the quality of predicted structures remains a bottleneck to the effectiveness of such models.
Recent efforts have explored the integration of multiple modalities of protein data to create more comprehensive representations. \cite{deepfri, heal, esm-gearnet, lm-gvp} introduce a joint protein representation for predicting protein functions by integrating the PLMs with graph-network-based structure encoders. In addition to deep mining of sequence and structural information, protein representation learning can also benefit from multi-modal approaches that integrate information from different sources, such as protein surface information \cite{protein-inr}, gene ontology annotation \cite{proteinssa, massa}, 3D point clouds \cite{nguyen2023multimodal} and sequence homology \cite{netgo, mulaxialgo}. By combining sequence, structure, and other data, these multi-modal representations offer a holistic view of protein characteristics, enabling more accurate predictions and deeper insights into protein functionality. 

\textbf{Protein domains.} 
Protein domains are structural units within proteins that play crucial roles in determining protein function. They can act alone or in concert with other domains to carry out the biological functions of the protein. Studying protein domains helps scientists better understand the function and structure of proteins, providing important insights for areas such as drug design and disease treatment. \cite{deepgraphgo, graph2go, sdn2go} integrate protein domain information into multi-modal features and achieve accurate predictions of GO terms. \cite{torres2021protein, msf-pfp, golabeler} employ a computational framework to fuse multi-source data features, including domains. However, these methods only use the category label of the domain without considering its correlation with protein function. 
\cite{domainpfp, dom2vec, domfun} learn associations between protein domains and functions combined at the protein level to derive functionally consistent representations for domains.
%\cite{domainpfp, dom2vec} employ self-supervised protocols to derive functionally consistent representations for domains. \cite{domainpfp, domfun} learn associations between protein domains and functions combined at the protein level and generate protein-function predictions. 
\cite{forslund2008predicting, drdo} develop new methods to infer protein functions based on protein domain combinations and domain order. Inspired by these methods, we propose to integrate domain information, including their functions and combinations, into a multimodal representation.

% ------------------------------------------------------------------------------------------
%建议加个Preliminary，包括PROBLEM STATEMENT、function priors计算、joint domain定义等
\section{Preliminaries}

\subsection{Problem formulation}

% 这个地方可以说明一下为什么选这个任务，为什么domain对这个任务比较有效果

Protein function prediction is a diverse and extensive field, covering topics like protein-ligand interactions and mutation effect prediction. Protein domains, being structural units made up of many residues, are particularly well-suited for function prediction tasks at the monomer level. In this work, we focus on a set of well-established protein function annotation benchmarks, all of which are consistently defined by maximize the likelihood:
\begin{equation}
\max_{\theta} P\left(y | x_{seq}, x_{str}, x_{dom}; \theta\right)
\end{equation}
where sequences are served as $x_{seq}$, structures are served as $x_{str}$, and domains are served as $x_{dom}$. $y$ represents the labels of protein functions. $\theta$ is the parameter of the network.

In a protein, the amino acids are linked by peptide bonds and form a chain. A protein sequence is denoted as $x_{seq}=(s_1,s_2,...,s_n)$, where $s_i$ is a residue at the $i_{th}$ position and $n$ is the sequence length. For protein structures, we use the coordinates of backbone, $i.e.$, $p_i\in R^{3}$. The protein structure is denoted as $x_{str}=(p_1,p_2,...,p_{3n})$. 
These modalities are considered explicit, as they exhibit relatively shallow features that require complex networks to fully explore their relationship with protein function. Domains, on the other hand, are considered an implicit modality as they are the direct determiner of protein functions and are composed of specific structures, $e.g.$, $d_1 = (p_i,...,p_j)$ (where $0 \leq i < j \leq 3n$). The domains are donated as $x_{dom} = (d_1, d_2, ..., d_t)$. $t$ is the domain number, which varies in different protein.

In our work, without loss of generality, we employ the InterPro entries \cite{interpro} to represent protein domains with a total of $a$. While the structure and function of protein domains are generally fixed, we utilize the entry index instead of the specific structure to represent the domains. 

\subsection{Function priors}
\label{ssec:func-priors}

% 这一节可以讲一讲那些只使用domain，而没有进行预训练的工作，讲一下这篇工作进行预训练的优势（能用于数据量较小的下游任务，并减少过拟合）。

In contrast to previous works \cite{deepgraphgo, graph2go, sdn2go} that directly input domain index into prediction networks, we integrate functional priors into the domain embeddings through pre-training. This strategy helps prevent overfitting in downstream tasks with limited data, as it allows the network to leverage richer domain representations rather than relying solely on raw indices. As a result, our multi-modal framework becomes more robust, which is validated by our ablation study as shown in Table~\ref{tab:ablation}.

\subsection{Joint domains}
\label{ssec:joint-dom}

Proteins are typically composed of multiple domains, and their functions often result from the coordinated action of several domains. However, not all domains in a protein contribute to a specific function. Based on this, we define "joint domains" as a combination of domains that collectively determine a protein's function, denoted as $d_{[i,j]}=[d_i,...,d_j]$.  In this paper, the joint domains are continuous, $i.e.$, $\forall i\leq k\leq j, d_k\in d_{[i,j]}$.

\section{Approach}

% Figure 2: overall architecture
\begin{figure}
\centering
\includegraphics[width=1\textwidth]{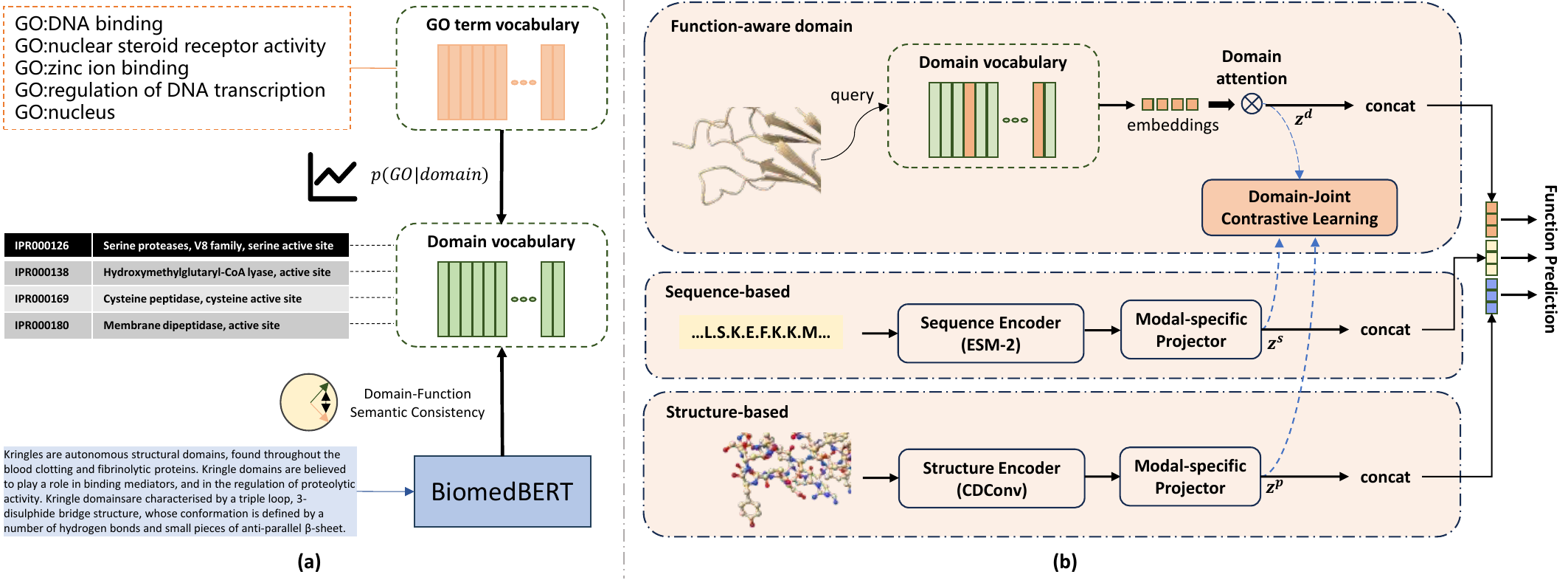}
\caption{
(a) \textbf{Function-aware domain embeddings pre-training.}
The domain vocabularies are pre-trained with function priors, including the domain-GO probabilities and the text semantic consistency.
(b) \textbf{Multi-modal function prediction architecture.}
We show how to augment sequence-based and structure-based models with function-aware domain and domain-joint contrastive learning.
}
\label{fig:overall}
\end{figure}

\subsection{Function-aware domain embeddings}
\label{ssec:fad}

% 将二值化特征使用embeddings表示，并说明将function的信息融入domain embeddings的思路
To incorporate the function priors into domain embeddings, we construct a dataset containing domain indexes, domain descriptions and GO terms.
Details of the construction steps can be found in Section~\ref{ssec:exp-setup}.
Building on this, we construct learnable vocabularies for both domain indexes and GO terms respectively as shown in Figure~\ref{fig:overall} (a). Then, we employ the domain-GO probabilities and the domain-text semantically consistent to update both vocabularies, together. 

\paragraph{Domain-GO probability}

GO terms are used in many bioinformatics tools and databases to help interpret and analyze experimental data, enabling researchers to gain insights into the functions of proteins. 
Before training the domain embeddings, we associate domains with the GO terms to mine the function priors of domains. Specifically, the domain indexes and GO terms can be represented as binary vectors $\{domain_i|domain_i\in [0,1]\}$ $(i=1,2,...,a)$ and $\{GO_j|GO_j\in [0,1]\}$ $(j=1,2,...,b)$, respectively, where $a,b$ are the vocabulary sizes of domain indexes and GO terms, respectively. $domain_i = 1$ indicates that a protein possesses the domain with the index of $i$ and $GO_j = 1$ indicates that a protein possesses the GO term with the index of $j$. 

Each protein independently possesses one or more domain indexes and GO terms. We define two types of associations. The association $(D_i, F_j, P_k)$ indicates that protein $P_k$ possesses the domain $i$ and the GO term $j$, while association $(D_i, P_k)$ indicates that protein $P_k$ possesses the domain $i$, regardless of the presence of GO terms.

We calculate the prior probability of the distribution of GO terms and utilize it to enhance the functional representation of domain vocabularies.
Specifically, the conditional probability of a protein that contains domain $i$ having the GO term $j$ is:
\begin{equation}
p\left(\text{GO}_j \mid \text{domain}_i\right)=\frac{p\left(\text{domain}_i, GO_j\right)}{p\left(\text{domain}_i\right)} = \frac{\sum_{k=1}^N I\left(D_i, F_j, P_k\right)}{\sum_{k=1}^N I\left(D_i, P_k\right)}
\end{equation}
where $N$ is the total number of protein samples and the operator $I(\cdot)$ indicates whether the specific association exists. The conditional probabilities serve as pseudo labels to train domain embeddings. Following DomainPFP \cite{domainpfp}, we adopt a simple network structure consisting of a Hadamard product operator and a few feedforward layers to encourage the domain vocabulary to learn functional relevance. Finally, a mse loss is employed to train the whole network. 

% 使用对比损失学习domain和function的语义一致性
\paragraph{Domain-text semantically consistent}

We further enrich the domain vocabulary with the functional information embedded in the textual descriptions. Specifically, we embed the descriptions by pre-trained BiomedBERT \cite{biomedbert} and train the domain vocabulary with contrastive learning: 
\begin{equation}
L_{sem}=-\log \frac{\exp \left(\operatorname{sim}\left(f(\mathbf{\phi}_i), f(\Phi(\mathbf{T}_i))\right) / \tau\right)}{\sum_{j=1}^N \exp \left(\operatorname{sim}\left(f(\mathbf{\phi}_i), f(\Phi(\mathbf{T}_j))\right) / \tau\right)}
\end{equation}
where $f(\cdot)$ is a learnable projector to project the embeddings into the shared semantic space, $\Phi$ represents the text encoder, $\phi_i\in \mathbf{R}^c$ represents the embedding of domain i, and $T_i$ is the textual description of domain i. 

The combination of two loss functions allows domain vocabularies and GO vocabularies to learn information that blends with each other. Subsequently, we employ the domain vocabularies, which are referred to as function-aware domain (FAD) embeddings, to augment existing sequence and structure-based representation.

\subsection{Architecture}
\label{ssec:framework}

We develop a multi-modal framework for function prediction, as shown in Figure~\ref{fig:overall} (b). The sequences and structures are processed using existing encoders, while the domain embeddings are obtained by retrieving them from the domain vocabulary. 
We employ a protein domain attention module (with the box positional encodings) to adaptively extract the functional representation of joint domains (more details are provided in the appendix). 
To adapt the features extracted by frozen modules for protein function prediction, we input the encodings of each modality into three separate modal-specific projectors and map them to feature vectors of the same dimension $c^*$. 
Through modal-specific projectors, we obtain the sequence encoding (served as $z^s \in \mathbf{R}^{c^*}$), the structure encoding (served as $z^p \in \mathbf{R}^{c^*}$), and the domain encoding (served as $z^d \in \mathbf{R}^{c^*}$). Finally, we concatenate the encodings $[z^s,z^p,z^d]$ as multi-modal features for protein function prediction.

In fact, without leveraging the sequence and structure modalities, the proposed function-aware domain embeddings alone can achieve efficient and robust function prediction, which is demonstrated in Table~\ref{tab:exp1} and Table~\ref{tab:exp2}. 

% ------------------------------------------------------------------------------------------
\subsection{Domain-joint contrastive learning}
\label{ssec:contrastive}

% Figure 3: domain-joint contrastive learning

To better align the various modalities in the representation space, we utilize the shared information of modalities for the same joint domains. We first introduce the semantic-enhanced embeddings for a more robust representation. Then, we develop a novel contrastive strategy (i.e. domain-joint contrastive learning) to overcome the noise of protein data. 

% 先介绍基本的对比学习，再提出motivation（蛋白质同源性引入噪声），进一步提出joint-domain可以产生不同的function，然后提出domain-joint contrastive learning
\paragraph{Inherent modality enhancement}

The semantics in the original input data are often complex, and some information is inevitably lost when encoding it into the feature space. When connecting and aligning existing representation spaces, this loss and bias of meaning will be inherited and amplified, affecting the robustness of alignment. Inspired by \cite{cmcr}, we add zero-mean Gaussian noise into the features and project them to the unit hyper-sphere with L2 normalization:
\begin{equation}
\begin{array}{lll}
\tilde{\mathbf{z}}^s = \operatorname{Normalize} \left( \mathbf{z}^s + \boldsymbol{\xi}_1 \right) ; & \tilde{\mathbf{z}}^p = \operatorname{Normalize} \left( \mathbf{z}^p + \boldsymbol{\xi}_2 \right) ; & \tilde{\mathbf{z}}^d = \operatorname{Normalize} \left( \mathbf{z}^d + \boldsymbol{\xi}_3 \right) ;
\end{array}
\label{eq:proj}
\end{equation}
where noise items $\xi_1, \xi_2, \xi_3 \in \mathbf{R}^{c^*}$ are sampled from zero-mean Gaussian distribution with variance $\sigma^2$, and they are not learnable. 
%Hence, aligning two embeddings with noise forces the model to acquire the ability to align all the embeddings within the two hyper-sphere. In the aligned space, the closer two embeddings are to each other, the more similar their semantics are. Embeddings within the same hyper-sphere share similar general semantics, and the semantics represented by the hyper-sphere are more comprehensive and robust than the original embedding.
Hence, aligning two embeddings with noise forces the model to acquire the ability to align all the embeddings within the two circles, leading to a more comprehensive and robust semantic representation.

\paragraph{Domain-joint alignment}

To establish the connection between two modalities, we project the semantic-enhanced embeddings (i.e. $\tilde{\mathbf{z}}^s, \tilde{\mathbf{z}}^p, \tilde{\mathbf{z}}^d$) to a new shared space \cite{cmcr, gmc} via a knowledge-shared projector $f(\cdot)$, respectively.
\begin{equation}
\hat{\mathbf{z}}^s = f\left( \tilde{\mathbf{z}}^s \right) ; \quad \hat{\mathbf{z}}^p = f\left( \tilde{\mathbf{z}}^p \right) ; \quad \hat{\mathbf{z}}^d = f\left( \tilde{\mathbf{z}}^d \right)
\end{equation}

In the projected space, our objective is to ensure that embeddings with similar semantics are close to each other. The various modalities $(x_{seq}, z_{str}, d_{dom})$ from the same protein is naturally semantically consistent, and it can be considered as a positive paired sample for contrastive learning. And different proteins are considered as negative samples in previous work \cite{gearnet, hermosilla2022contrastive}. 
% 因为存在负样本噪声，我们提出使用domain-joint来解决样本噪声问题，将对比损失转换为domain-joint contrastive，这里可以补充一下使用domain-joint的理论依据（就是不同的joint domains会决定不同的功能）
% 对应的补充材料中推导出我们给的公式是对比损失的一个特殊形式，并且使用信息论证明该损失也是增大互信息理论下限。
However, there are function-similar proteins ($e.g.$ homologous proteins) that introduce noise to the vanilla contrastive training. As discussed in \cite{albef}, the repulsive structure in contrastive loss penalizes all data pairs, regardless of the potential correlations between negative samples. 

Inspired by recent advancements in contrastive learning methods \cite{chen2020simple}, we propose to perform domain-joint cropping on proteins and construct negative samples with various sub-views to ensure feature discriminability between positive and negative samples. Specifically, we randomly sample various joint domains while ensuring that there is no overlap between them. In the experiment, we utilize two sets of joint domains, $\zeta_1(x_i)$ and $\zeta_2(x_i)$. The sequence-domain contrastive loss $L_{sdc}$ and structure-domain contrastive loss $L_{pdc}$ are defined as (more details are provided in the Appendix):
\begin{equation}
L_{sdc} = \sum_i^N \left[ \underbrace{-\operatorname{sim}\left(\hat{\mathbf{z}}^s(\zeta_1(x_i)), \hat{\mathbf{z}}^d(\zeta_1(x_i))\right)}_{L_{imc}^s\text{: pull positive close}} + \underbrace{\operatorname{sim}\left(\hat{\mathbf{z}}^s(\zeta_1(x_i)), \hat{\mathbf{z}}^d(\zeta_2(x_i))\right)}_{L_{ivd}^s\text{: push negative away}} \right]
\label{eq:sdc}
\end{equation}
\begin{equation}
L_{pdc} = \sum_i^N \left[ \underbrace{-\operatorname{sim}\left(\hat{\mathbf{z}}^p(\zeta_1(x_i)), \hat{\mathbf{z}}^d(\zeta_1(x_i))\right)}_{L_{imc}^p\text{: pull positive close}} + \underbrace{\operatorname{sim}\left(\hat{\mathbf{z}}^p(\zeta_1(x_i)), \hat{\mathbf{z}}^d(\zeta_2(x_i))\right)}_{L_{ivd}^p\text{: push negative away}} \right]
\label{eq:pdc}
\end{equation}
where $x_i$ represent a complete protein, $\zeta_1, \zeta_2$ represent various sub-views divided according to the joint domains and $N$ is the number of proteins. 
The first term is the inter-modality consistency loss which enhances the semantic consistency between multi-modal representations. The second term is the inter-view distinctiveness loss which encourages the representation to efficiently distinguish different protein functions. 
% 存在模态gap问题，MCR通过去除第二项解决，但是会导致互信息降低
% 对应的补充材料使用图示解释模态gap的成因和互信息降低
However, the analysis in \cite{liang2022mind} suggests that different data modalities are embedded at arm's length in their shared representation in multi-modal models, which is termed as modality gap. It is demonstrated that contrastive learning keeps the different modalities separated by a certain distance and varying the modality gap distance has a significant impact on improving the model's downstream zero-shot classification performance and fairness. 

\cite{cmcr} proposes closing the modality gap and guaranteeing that embeddings from different modalities with similar semantics are distributed in the same region of the representation space by removing the repulsive structure in the contrastive loss. However, simply deleting the repulsive structure easily leads to reducing the mutual information (MI) between modalities and cannot keep task-relevant information intact, which leads to decreasing downstream classification accuracy as discussed in \cite{tian2020makes}. 

% 提出新的三元组损失来替换push项，有效地增大了互信息
%（补充材料使用图示解释增大样本gap，可以证明提高了互信息下限?）
Inspired by \cite{schroff2015facenet}, we construct triplets with various modalities of the sub-views and propose a triplet loss to replace the inter-view distinctiveness loss in Equation~\ref{eq:sdc} (same for Equation~\ref{eq:pdc}). Specifically, we use the structure of one of the sub-views as the anchor, the joint domains of that sub-view as the positive sample, and the joint domains of the other sub-view as the negative sample. The newly inter-view distinctiveness loss can be formulated as:
\begin{equation}
L_{ivd}^p = \sum_i^N \left[ -\operatorname{sim}\left(\hat{\mathbf{z}}^p(\zeta_1(x_i)), \hat{\mathbf{z}}^d(\zeta_1(x_i))\right) + \operatorname{sim}\left(\hat{\mathbf{z}}^p(\zeta_1(x_i)), \hat{\mathbf{z}}^d(\zeta_2(x_i))\right) + \alpha\right]_{+}
\end{equation}
where $\alpha$ is a hyper-parameter employed to control the gap between protein representations from different sub-views (more details are provided in the Appendix). We combine the inter-modality consistency loss and the new inter-view distinctiveness, which is referred to as the triplet InfoNCE loss, to close the modality gap while maintaining the differences between different sub-views.
\begin{equation}
    L_{triplet}^p = L_{imc}^p + \lambda L_{ivd}^p
\end{equation}
where $\lambda$ is a hyper-parameter that adjusts training stability. When $\lambda$ is small, it is less likely to create modality gaps, but training efficiency decreases. 
Finally, the domain-joint contrastive learning is supervised by the combination of the sequence-domain triplet loss $L_{triplet}^s$ and the structure-domain triplet loss $L_{triplet}^p$.
%By aligning domain-guided cross-modal semantically consistent embeddings in the knowledge-shared space, i.e., aligning $(\hat{z_i^s}, \hat{z_i^d})$ and $(\hat{z_i^p}, \hat{z_i^d})$, the embeddings of various modalities are semantically consistent and able to distinguish different protein functions. 

%effectively increasing the mutual information between different modalities. More details are provided in the Appendix.

%As shown in Figure \ref{fig:pc-mcr}, the inter-modality consistency loss encourages all features from the same protein to be projected onto the same circle in the PC-MCR space, while the inter-view distinctiveness loss attempts to mine the distinctiveness between different domain combinations within each protein.

%\begin{figure}[htbp]
%\centering
%\includegraphics[width=0.8\textwidth]{figure/PC-MCR.png}
%\caption{Inherent modality alignment enhances the semantic consistency between multi-modal representations, while protein cropping alignment amplify the gap between different domain representations.}
%\label{fig:pc-mcr}
%\end{figure}

% ------------------------------------------------------------------------------------------
\section{Experiments}
\label{sec:exp}

% 这里后面有几篇新工作可以加一下

\begin{table}
\caption{Fmax of gene ontology term prediction and enzyme commission number prediction. *Results are from \cite{cdconv}. †Results are from \cite{gearnet}. **Results are from \cite{esm-gearnet}}
\centering
\begin{tabular}{ccccccc}
%\begin{tabular}{c|c|c|cccc}
    \toprule
    \multirow{2}{*}{Input} & \multirow{2}{*}{Method} & Additional & \multicolumn{3}{c}{Gene Ontology} & Enzyme \\
    \cmidrule(r){4-6}
     &  & Modality & BP & MF & CC & Commission \\
    \midrule
    \multirow{3}{*}{Sequence} & ESM-1b\cite{esm1b}† & - & 0.452 & 0.657 & 0.477 & 0.864 \\
     & \textbf{ESM-2}\cite{esm2}** & - & 0.460 & 0.661 & 0.445 & 0.880 \\
     & SaProt\cite{saprot} & - & 0.356 & 0.678 & 0.414 & 0.884 \\
    \midrule
    \multirow{4}{*}{Structure} & GVP\cite{gvp}* & - & 0.326 & 0.426 & 0.420 & 0.489 \\
     & IEConv\cite{ieconv}* & - & 0.421 & 0.624 & 0.431 & 0.735† \\  
     & GearNet\cite{gearnet}† & - & 0.356 & 0.503 & 0.414 & 0.730 \\
     & \textbf{CDconv}\cite{cdconv}* & - & 0.453 & 0.654 & 0.479 & 0.820 \\
    \midrule
    Sequence & DeepFRI\cite{deepfri}† & - & 0.399 & 0.465 & 0.460 & 0.631 \\
    \& & LM-GVP\cite{lm-gvp}† & - & 0.417 & 0.545 & 0.527 & 0.664 \\
    Structure & ESM-GearNet\cite{esm-gearnet}** & - & 0.488 & 0.681 & 0.464 & \underline{0.890} \\   
    \midrule
    \multirow{4}{*}{Multimodal} & ProteinINR\cite{protein-inr} & surface & 0.508 & 0.678 & 0.506 & 0.890 \\ %Multimodal
    & ProteinSSA\cite{proteinssa} & GO terms & 0.464 & 0.667  & 0.492  & 0.857 \\
    \cmidrule(r){2-7}
     & \textbf{FAD} (only) & domain & \underline{0.511} & \underline{0.698} & \underline{0.533} & 0.878 \\
     %& \textbf{CDConv-FAD} & domain & 0.500 & 0.694 & 0.537 & \textbf{0.907} \\
     & \textbf{ProtFAD} & domain & \textbf{0.518} & \textbf{0.701} & \textbf{0.551} & \textbf{0.911} \\
    \bottomrule
\end{tabular}
\label{tab:exp1}
\end{table}

\begin{table}
\centering
\caption{Accuracy of protein fold classification and enzyme catalytic reaction classification. *Results are from \cite{cdconv}. †Results are from \cite{gearnet}.}
\begin{tabular}{cccccc}
%\begin{tabular}{c|c|c|ccc}
    \toprule
    \multirow{2}{*}{Input} & \multirow{2}{*}{Method} & Additional & \multicolumn{2}{c}{Fold Classification} & Enzyme \\
    \cmidrule(r){4-5}
     &  & Modality & Superfamily & Family & Reaction \\
    \midrule
    \multirow{2}{*}{Sequence} & ESM-1b\cite{esm1b}† & - & 0.601 & 0.978 & 0.831 \\
     & \textbf{ESM-2}\cite{esm2} & - & 0.789 & 0.992 & 0.894 \\
     %& ESM-2\cite{esm2} & - &  &  &  &  \\
     %& SaProt\cite{saprot} & - &  &  &  &  \\
    \midrule
    \multirow{4}{*}{Structure} & GVP\cite{gvp}* & - & 0.225 & 0.838 & 0.655 \\
     & IEConv\cite{ieconv}* & - & 0.702 & 0.992 & 0.872 \\ 
     & GearNet\cite{gearnet}† & - & 0.805 & \textbf{0.999} & 0.875 \\
     & \textbf{CDconv}\cite{cdconv}* & - & 0.777 & 0.996 & 0.885 \\
    \midrule
    \multirow{1}{*}{Sequence \& Structure} & DeepFRI\cite{deepfri}† & - & 0.206 & 0.732 & 0.633 \\
     %& LM-GVP\cite{lm-gvp} & - &  &  &  &  \\
     %& ESM-GearNet\cite{esm-gearnet} & - &  &  &  &  \\
    \midrule
    \multirow{3}{*}{Multimodal} %& ProteinINR\cite{protein-inr} & surface &  &  &  &  \\ %Multimodal
    & ProteinSSA\cite{proteinssa} & GO terms & 0.794 & 0.998 & \underline{0.894} \\
    \cmidrule(r){2-6}
     & \textbf{FAD} (only) & domain & \underline{0.850} & 0.991 & 0.867 \\
     %& \textbf{CDConv-FAD} & domain &  &  &  &  \\
     & \textbf{ProtFAD} & domain & \textbf{0.908} & \underline{0.998} & \textbf{0.923} \\
    \bottomrule
\end{tabular}
\label{tab:exp2}
\end{table}

\subsection{Experimental setups}
\label{ssec:exp-setup}

\textbf{Domain embeddings pre-training.}
%\textbf{Dataset of domains and GO annotations.}
We collected 570,830 protein entries from Swiss-Prot \cite{uniprot} (release 2024\_1) including the InterPro IDs and GO term IDs. We discarded all proteins which had no InterPro annotations. We also collected the mappings of InterPro entries to GO terms and textual descriptions from the InterPro Database \cite{interpro} (release 2023\_10). In summary, our dataset contained 551,756 proteins with 31,929 unique domains and 28,944 unique GO terms. As introduced in Section \ref{ssec:fad}, We finally retained 1,454,811 domain-GO paired samples. We pre-train the FAD embeddings for 500 epochs with a batch size of 16,384. We set the embedding dimension to 768 to march sequence encodings with a dimension of 1280 and structure encodings with a dimension of 2048. After training, we freeze the parameters of the FAD embeddings for downstream tasks.

\textbf{Benchmark tasks.} Following \cite{ieconv, gearnet, cdconv}, we evaluate the proposed method on four tasks: protein fold classification \cite{hermosilla2022contrastive, ieconv}, enzyme reaction classification \cite{ieconv}, gene ontology (GO) term prediction \cite{deepfri} and enzyme commission (EC) number prediction \cite{deepfri}. 
Protein fold classification includes two evaluation scenarios: superfamily and family (the fold scenario in \cite{cdconv} is not relevant to protein function prediction, so we removed it). GO term prediction includes three sub-tasks: biological process (BP), molecular function (MF), and cellular component (CC) ontology term prediction. Note that we use domains instead of GO terms as input, so there is no data leakage.
For each protein in the datasets, we assigned InterPro domains using InterProScan 5 \cite{interproscan}. 
Protein fold and enzyme reaction classification are single-label classification tasks. Mean accuracy is used as the evaluation metric. GO term and EC number prediction are multi-label classification tasks. The Fmax accuracy is used as the evaluation metric.

%More details about these tasks and datasets are shown in Appendix X. 

\textbf{Implementation.} In our evaluation, we employ pre-trained ESM-2 \cite{esm2} as the sequence encoder and CDConv \cite{cdconv} as the structure encoder. The parameters of ESM-2 are frozen to reduce unnecessary training overhead. We adopt simple multi-layer perceptrons as our projectors. The variance $\sigma^2$ of the noises in Equation \ref{eq:proj} is set as 0.01. We train our model using the same configuration as in \cite{cdconv}.

\begin{figure}
\centering
\includegraphics[width=1\textwidth]{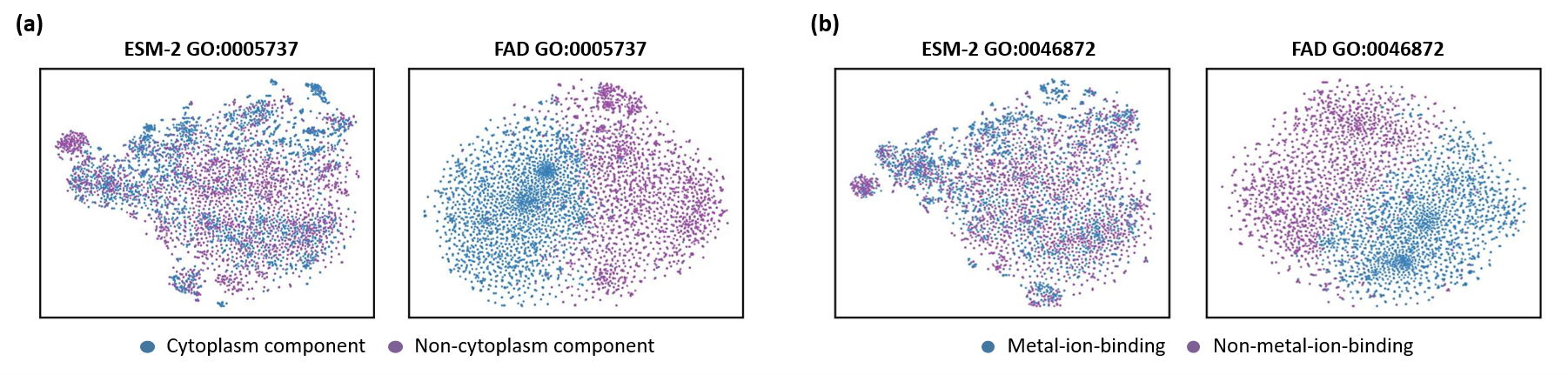}
\caption{
The dimensionality-reduction visualizations of domain embeddings extracted by ESM-2 and our function-aware domain (FAD).
Two common types of functions, including "cytoplasm" (a) and "metal-ion-binding" (b), are selected.
%Blue points are samples with the specific function, while purple points are corresponding negative samples.
}
\label{fig:visualize}
\end{figure}

\begin{table}
\caption{Ablation study of the proposed modules.}%Fmax is used for gene ontology term prediction (GO) and enzyme commission number prediction (EC). Accuracy is used for protein fold classification (FC) and enzyme catalytic reaction classification (ER).}
\centering
\begin{tabular}{cccccccc}
%\begin{tabular}{c|ccccccc}
    \toprule
    \multirow{2}{*}{Method} & \multicolumn{2}{c}{FC} & \multirow{2}{*}{ER} & \multicolumn{3}{c}{GO} & \multirow{2}{*}{EC} \\
    \cmidrule(r){2-3}
    \cmidrule(r){5-7}
     & Superfamily & Family &  & BP & MF & CC &  \\
    \midrule
    \textbf{ProtFAD} & \textbf{0.908} & \textbf{0.998} & \textbf{0.923} & \textbf{0.518} & \textbf{0.701} & \textbf{0.551} & \textbf{0.911} \\
    \midrule
    w/o domain & 0.830 & 0.995 & 0.906 & 0.485 & 0.672 & 0.519 & 0.875 \\	
    w/o domain pre-train & 0.881 & 0.996 & 0.909 & 0.495 & 0.683 & 0.532 & 0.868 \\
    w/o domain attention & 0.871 & 0.997 & 0.909 & 0.508 & 0.697 & \underline{0.545} & 0.899 \\
    w/o contrastive loss & \underline{0.902} & 0.997 & 0.910 & \underline{0.514} & \underline{0.699} & 0.528 & 0.904 \\
    w/ vanilla contrastive loss & 0.891 & \underline{0.997} & \underline{0.911} & 0.512 & 0.698 & 0.520 & \underline{0.905} \\
    \bottomrule
\end{tabular}
\label{tab:ablation}
\end{table}

\subsection{Comparison with state-of-the-art}

We compare our method with existing sequence-only, structure-only, and multi-modal methods. Results are shown in Table \ref{tab:exp1} and Table \ref{tab:exp2}. To distinguish between methods with and without the inclusion of a third modality, we denote methods that only utilize sequence and structure as "Sequence \& Structure" and methods that utilize more modalities as "Multimodal" with a description of the third modality. Additionally, we present results for our method using only domain modality (denoted as "FAD") and the method using information from sequence, structure, and domain(denoted as "ProtFAD"), to demonstrate that the effectiveness of the proposed FAD and multi-modal representations incorporating FAD. Note that our ProtFAD significantly outperforms all the existing methods on nearly all benchmarks. 
Even if only the domain modality is used, FAD shows competitive results, which showcases the effectiveness of introducing function-aware domains as implicit modality.

\subsection{Visualization}

We employ t-SNE to visualize the domain representations generated by FAD and ESM-2. We select two common types of functions, $i.e.$ "cytoplasm" and "metal-ion-binding". For each function, we sample an approximately equal number of positive and negative domains. We embed the domains by FAD and embed the sequences corresponding to the domains by ESM-2. 
%Note that a domain may correspond to multiple sequences, and we randomly select one of them. 
As shown in Figure \ref{fig:visualize}, the representations generated by ESM-2 are intertwined, whereas those generated by FAD are clearly separated between the positive and negative samples, demonstrating that FAD accurately captures intricate protein functions.

\subsection{Ablation study}

To analyze the effect of various components, we perform an ablation study on four tasks and present results in Table \ref{tab:ablation}. We first examine two degenerate settings of function-aware domain embeddings, i.e., "w/o domain" and "w/o domain pre-train". 
"w/o domain pre-train" indicates that we use binary features to represent domains, rather than using pre-trained embeddings. 
These settings lead to a deterioration in performance across all benchmark tasks, highlighting the importance of domain with function priors for protein function prediction. 
Next, we remove the domain attention mechanism and use the mean of domain embeddings instead, resulting in consistent performance degradation too.
Besides, we investigate the impact of excluding the protein domain-joint contrastive learning from our model architecture. The declining results further demonstrate the individual contributions of the proposed modules, as evidenced in the table.

% 模态特征提取inbalance，过度focus在FAD部分，整体可能没到最优，Fold数据集崩坏，未来工作可以优化

% ------------------------------------------------------------------------------------------
\section{Conclusion}
\label{sec:con}

In this work, we propose ProtFAD, a priors-guided multi-modal protein representation learning approach, to bridge the gap between sequence or structure modality and protein functions.
By leveraging function-aware domain embeddings and a novel domain-joint contrastive learning, we extract rich functional information from protein domains and enhance existing protein representation. During the fusion of multi-modal features, we further incorporate domain positional information and the synergistic effects between domains with a domain attention mechanism. Extensive experiments on diverse protein function prediction benchmarks verify the superior performance of ProtFAD.

{
\small
\bibliographystyle{unsrt}  
\bibliography{references}
}

%%%%%%%%%%%%%%%%%%%%%%%%%%%%%%%%%%%%%%%%%%%%%%%%%%%%%%%%%%%%

\appendix

\section{Additional modules}

\subsection{Domain-protein-GO term association}

\begin{figure}[htbp]
\centering
\includegraphics[width=0.6\textwidth]{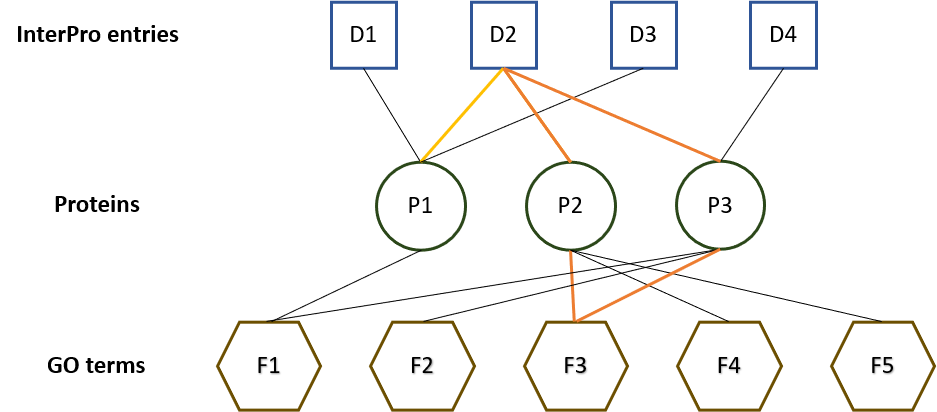}
\caption{Establish associations between domains and GO terms through proteins. Define two distinct meta-paths and utilize them to summarize the connection between domains and GO terms, including domain-protein-function (red line) and domain-protein (yellow line).}
\label{fig:connect-graph}
\end{figure}

Here, we establish the association between domains and GO terms. We collect the proteins from Swiss-Prot \cite{uniprot} including the InterPro IDs and GO term IDs, and establish a relationship network among them as shown in Figure \ref{fig:connect-graph}. When a protein k containing domain i has the GO j, we define a meta-path $(D_i, F_j, P_k)$. When a protein k contains the domain i, we define another meta-path $(D_i, P_k)$. Two meta-paths may overlap for some proteins. The meta-paths are referred to as associations in Section \ref{ssec:func-priors} and are used to calculate the function priors in Section \ref{ssec:fad}.

\subsection{Domain attention module}
\label{sec:domain-attention}

% 说明蛋白质的功能是多个domains共同作用的结果
Protein function may be determined by several domains, and different domains may contribute to different functions. Therefore, we employ a protein domain attention module to adaptively extract the functional representation of joint domains.
% 详细介绍提出的domain attention模块的组成
Specifically, we use $\mathbf{e} = [e_1, e_2, ..., e_t]$ to represent the FAD embeddings of joint domains $d_{[1,t]}$. We adopt a self-attention layer to calculate the importance of each domain, as shown below:
\begin{equation}
\hat{d} = \textbf{Agg}(\mathbf{e} + \Omega(\mathbf{e}))
\end{equation}
where $\Omega(\cdot)$ is the self-attention layer and $\textbf{Agg}(\cdot)$ is an aggregation operator (which is average pooling in our experiment).

% 融入domain的位置信息
Considering that the position of the domain may affect the function of the protein, we incorporate a positional encoding for the domain. 
The sequence length and relative position of each domain within a protein are different. For simplicity, we take the position of the amino acid in the middle of the domain sequence as the position of the domain and normalize it to (0, 1) by dividing it by the length of the domain sequence.
% 这里介绍box position encoding
Common discrete positional encodings cannot be used for the continuous position values described above. Therefore, we employ box positional encodings. 
Specifically, we group continuous position $p$ within the interval (0, 1) into bins and learn a unique position embedding $\psi$ for each bin. Finally, the calculation of domain attention can be represented as:
\begin{equation}
\hat{e}_i = e_i + \psi_{\lfloor b\times p_i \rfloor}, \quad i=1,2,...,t
\end{equation}
\begin{equation}
\hat{d} = \textbf{Agg}(\mathbf{\hat{e}} + \Omega(\mathbf{\hat{e}}))
\end{equation}
where $b$ is the number of bins, and $\psi_j$ represents the position embedding of $j_{th}$ bin.

This attention module considers the relationships between different domains and the positional information of each domain and generates a single domain-joint representation $\hat{d} \in \mathbf{R}^c$, enabling the model to effectively capture the complex interplay between domains and their contributions to protein function.

In addition, we explore the performance of different positional encodings (PE):

(1) BERT positional encoding: learn a position embedding for each element in joint domains ($i.e. i=1,2,...,t$). This approach considers the relative positional relationship between domains without receiving their real positions as input.

(2) Field positional encoding: learn a position embedding for all domains, and the positional encoding is the product of the embedding and the domain position $p$. This approach integrates the positional information with a linear function.

(3) MLP positional encoding: employ one MLP projecting the domain position $p$ to a position vector for each domain.

\begin{table}
\caption{Comparative study of various positional encodings (PE) on enzyme commission number prediction (EC). The methods are evaluated in the ProtFAD framework without domain-joint contrastive learning.}
\centering
\begin{tabular}{cccccc}
    \toprule
    positional encoding & w/o PE & BERT PE & Field PE & MLP PE & Box PE \\
    \midrule
    enzyme commission & 0.9024 & 0.8998 & 0.9015 & \underline{0.9045} & \textbf{0.9050} \\
    \bottomrule
\end{tabular}
\label{tab:pos-encoding}
\end{table}

The results are shown in Table \ref{tab:pos-encoding}. The box position encoding achieves the best performance. Note that BERT PE and Field PE are less effective than not using the positional encoding, proving that the continuous position $p$ cannot be modeled with common discrete encodings.

\section{Proof}
\label{app:proof}

\subsection{Derivation of Equation \ref{eq:pdc}}

We perform domain-joint cropping on proteins and construct various sub-views, i.e. $\zeta_1(x), \zeta_2(x), ..., \zeta_K(x)$. We consider the multiple modalities of the same sub-view as positive samples, and the different sub-views as negative samples. We utilize two modalities for contrastive learning. For structure and domain, the contrastive loss is:
\begin{equation}
L_{pdc} = -\log \frac{\exp \left(\operatorname{sim}\left(\hat{\mathbf{z}}^p(\zeta_1(x)), \hat{\mathbf{z}}^d(\zeta_1(x))\right) \right)}{\sum_{k\neq 1}^K \exp \left(\operatorname{sim}\left(\hat{\mathbf{z}}^p(\zeta_1(x)), \hat{\mathbf{z}}^d(\zeta_k(x))\right) \right)}
\end{equation}
where K is the number of sub-views, $\zeta_k$ represents the k-th protein sub-view. Let K=2, we get
\begin{equation}
L_{pdc} = -\operatorname{sim}\left(\hat{\mathbf{z}}^p(\zeta_1(x)), \hat{\mathbf{z}}^d(\zeta_1(x))\right) + \operatorname{sim}\left(\hat{\mathbf{z}}^p(\zeta_1(x)), \hat{\mathbf{z}}^d(\zeta_2(x))\right)
\end{equation}

%\paragraph{Mutual Information}

%\begin{equation}
%f_k(a,b) \propto \frac{p(a,b)}{p(a)p(b)}
%\end{equation}

%\begin{equation}
%L_N = - E_X [log \frac{f_k(a_1,b_1)}{f_k(a_1,b_2)}]
%\end{equation}

%\begin{align*}
%L_N^{opt} &= - E_X log \frac{\frac{p(a_1,b_1)}{p(a_1)p(b_1)}}{\frac{p(a_1,b_2)}{p(a_1)p(b_2)}} \\
%&= E_X log \frac{p(a_1)p(b_1)}{p(a_1,b_1)} \cdot E \frac{p(a_1,b_2)}{p(a_1)p(b_2)} \\
%&= E_X log \frac{p(a_1)p(b_1)}{p(a_1,b_1)} \\
%&= - I(a_1,b_1) \\
%\end{align*}

\subsection{Triplet InfoNCE Loss}

\begin{figure}[htbp]
\centering
\includegraphics[width=0.8\textwidth]{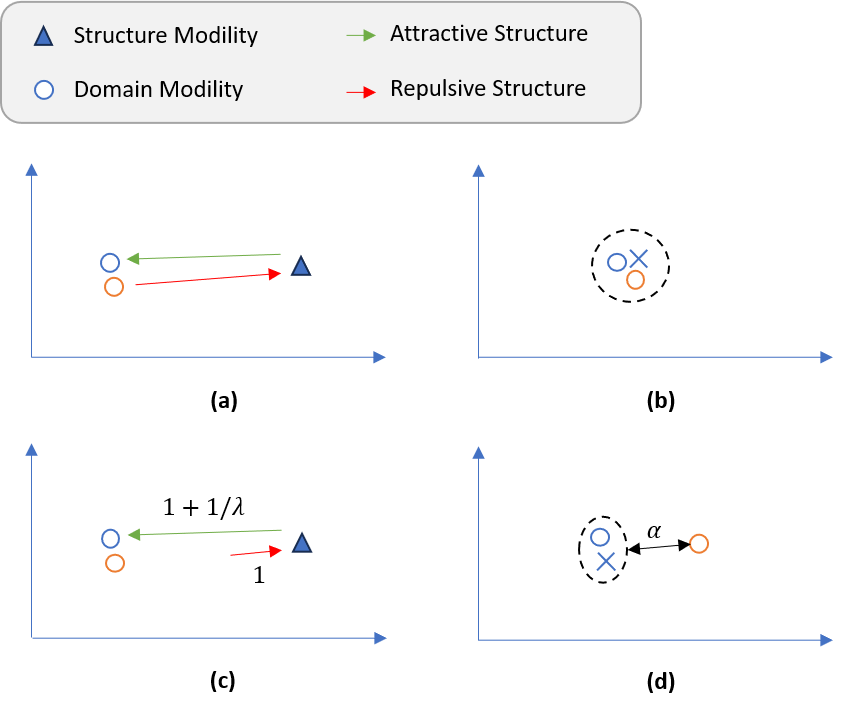}
\caption{(a) Attractive structure and repulsive structure jointly keep the modality gap. (b) After deleting the repulsive structure, the different samples cannot be distinguished (the dashed line represents the aggregated embeddings). (c) The triplet InfoNCE loss eliminates the modality gap by increasing the weight of the attractive structure. (d) The triplet InfoNCE loss enlarges the mutual information by maintaining the lower bound of sample distance.}
\label{fig:tri-info}
\end{figure}

The analysis in \cite{liang2022mind} suggests that contrastive learning keeps the different modalities separated by a certain distance. Here, we demonstrate how our proposed triplet InfoNCE loss closes the modality gap while keeping the sample distinctiveness. We denote two different samples as $z_1,z_2$, and use superscripts $p,d$ to indicate two different modalities (e.g. structure and domain). The contrastive loss of two sub-views can be expressed as:
\begin{equation}
L = \underbrace{d\left(z_1^p, z_1^d\right)}_{\text{attractive structure}} - \underbrace{d\left(z_1^p, z_2^d\right)}_{\text{repulsive structure}}
\end{equation}
where $d(A,B) = 1 - \operatorname{sim}\left(A, B\right)$ represent the cosine distance. Euclidean distance and cosine distance are monotonically related ($A,B$ are L2-normalized), that is $d_{Euclidean} = \sqrt{2d}$. So we use Euclidean distance to demonstrate the following process on a two-dimensional plane.

For example, the repulsive structure in the contrastive loss keeps the modality gap when similar samples are aligned as shown in Figure \ref{fig:tri-info} (a). Simply deleting the repulsive structure like \cite{cmcr} easily leads to aggregating embeddings with different semantics as shown in Figure \ref{fig:tri-info} (b), which reduces the mutual information (MI) between modalities.

For the triplet InfoNCE loss, when the loss is not converged, we get:
\begin{align*}
L_{triplet} &= d\left(z_1^p, z_1^d\right) - 1 + \lambda \left[ d\left(z_1^p, z_1^d\right) - d\left(z_1^p, z_2^d\right) + \alpha\right]_{+} \\
&= (1+\lambda) d\left(z_1^p, z_1^d\right) - \lambda d\left(z_1^p, z_2^d\right) -1 + \lambda\alpha \\
&= \lambda \left[ (1+\frac{1}{\lambda})d\left(z_1^p, z_1^d\right) - d\left(z_1^p, z_2^d\right) + \alpha - \frac{1}{\lambda} \right]
\end{align*}
where $\lambda$ is a hyper-parameter that adjusts training stability, and $\alpha$ is a hyper-parameter employed to control the gap between different samples. When $\lambda \to \infty$, $L_{triplet}$ will degenerate into a vanilla contrastive loss. And when $\lambda$ is small, it is less likely to create modality gaps as the repulsive structure accounts for less in the loss function, as shown in Figure \ref{fig:tri-info} (c).

When the loss is converged, we get:
\begin{equation}
d\left(z_1^p, z_1^d\right) - d\left(z_1^p, z_2^d\right) + \alpha \leq 0
\end{equation}

When the modalities are aligned, we expect $d\left(z_1^p, z_1^d\right) = 0$. That is
\begin{equation}
0 = d\left(z_1^p, z_1^d\right) \leq d\left(z_1^p, z_2^d\right) - \alpha
\end{equation}

$z_1,z_2$ are interchangeable, so we have
\begin{equation}
\begin{array}{ll}
d\left(z_1^p, z_2^d\right) \geq \alpha; & d\left(z_2^p, z_1^d\right) \geq \alpha
\end{array}
\end{equation}
\begin{equation}
\begin{array}{ll}
d\left(z_1^d, z_2^d\right) \geq \alpha; & d\left(z_2^p, z_1^p\right) \geq \alpha
\end{array}
\end{equation}
The $\alpha$ is the lower bound of the distance between different samples as shown in Figure \ref{fig:tri-info} (d), which keeps the mutual information between different modalities for better alignment \cite{tian2020makes, wang2020understanding}.

% 是否可以证明三元组损失通过保持样本gap提高了互信息下限

\section{More experiment details}
\label{app:experiment}

\subsection{Experiments compute resources}

We use Tesla V100-SXM2-32GB GPU for single-card training.

\section{Discussion}
\label{app:discussion}

\subsection{Domain embedding dimension}

\begin{figure}
\centering
\includegraphics[width=0.8\textwidth]{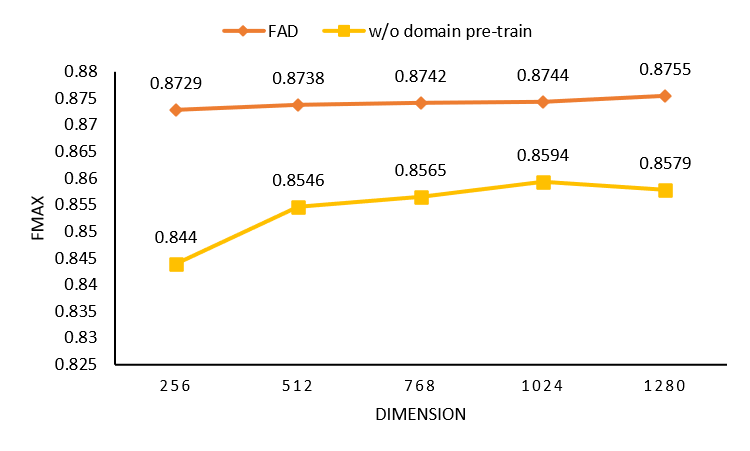}
\caption{Comparative study of various domain embedding dimensions on enzyme commission number prediction.}
\label{fig:dimension}
\end{figure}

To explore the density of information contained in the domain, we compare the impact of different domain embedding dimensions on model performance. Specifically, we only employ the domain modality for enzyme commission number prediction, comparing the results of FAD and domain embeddings without pre-training under different dimension settings. The results are shown in Figure \ref{fig:dimension}. 

As the dimension of FAD embeddings increases, the model performance maintains an improving trend, indicating that FAD does integrate effective functional information. In addition, it implies that FAD has greater potential for function perception as the dimension increases. For domain embeddings that are not pre-trained, the model performance reaches saturation when the embedding dimension is 1024. This may be caused by insufficient data in the function prediction task to train a stronger domain representation, i.e., the model is overfitting. This further illustrates the importance of pre-training our functional representations.

\subsection{Is sequence or structure necessary?}

We further evaluate the performance of ProtFAD without sequence modality or without structure modality, and provide the results in Table \ref{tab:sequence-compare}. The sequence modality contributes to the multi-modal representations for most benchmarks, especially the cellular component ontology term prediction (an improvement of nearly 10\% compared to the degenerate model). Surprisingly, adding sequence modality reduces the performance of enzyme commission number prediction. This may be caused as the information in structure and domain is sufficient for the task. However, for structure modality, the performance degradation is not obvious, which introduces that the information in domains is sufficient for the protein function prediction or the information mining in previous structure-based work is inadequate. It proves the correctness of using domains as an implicit modality to connect structure and function.
%It indicates that for specific tasks, the modalities used need to be specifically designed.

\begin{table}
\caption{Absolute study of ProtFAD without sequence modality on gene ontology term prediction and enzyme commission number prediction.}
\centering
\begin{tabular}{ccccc}
    \toprule
    \multirow{2}{*}{Method} & \multicolumn{3}{c}{gene ontology} & \multirow{2}{*}{enzyme commission} \\
    \cmidrule(r){2-4}
     & BP & MF & CC &  \\
    \midrule
    \textbf{ProtFAD} & \textbf{0.515} & \textbf{0.701} & \textbf{0.552} & \underline{0.907} \\
    \midrule
    w/o sequence & 0.500 & 0.694 & 0.496 & \textbf{0.909} \\	
    $\Delta$ & -1.5\% & -0.7\% & -5.6\% & +0.2\% \\
    \midrule
    w/o structure & \underline{0.515} & \underline{0.701} & \underline{0.548} & 0.906 \\	
    $\Delta$ & 0\% & 0\% & -0.4\% & -0.1\% \\
    \bottomrule
\end{tabular}
\label{tab:sequence-compare}
\end{table}

\subsection{How to crop the protein}
% 介绍joint domain的具体细节

In the experiment, we divide the domains of a protein ($i.e. x_{dom}=(d_1,d_2,...,d_t)$) into two subsets $\zeta_1(x_{dom})$ and $\zeta_2(x_{dom})$. Specifically, we random select $k\in(1,t]$, let $\zeta_1(x_{dom})=(d_1,...,d_{k-1})$, $\zeta_2(x_{dom})=(d_k,...,d_t)$. Then we search the corresponding sequence and structure of the divided domains for the two modalities. 

This division cannot separate the sequence and structure of the two sub-views, introducing noise into the contrastive learning. Reducing such noise can further enhance the effect of contrastive learning. However, we have not conducted an in-depth exploration of how domains are divided, which may be a meaningful future work.

\subsection{Limitation}

In the context of protein function prediction, there are some limitations when using domain-based approaches. First, since domains are structural units composed of multiple atoms or residues,enhancing function prediction based on protein domains in more granular tasks (such as binding site prediction) requires increasingly complex network architectures. This added complexity can introduce challenges in model design and optimization. Second, due to the inductive bias inherent to protein domains, domain-based function prediction may suffer from severe overfitting in cases where data is limited, despite improvements from function-prior pretraining. While such pretraining can mitigate overfitting to some extent, the fixed nature of domain-specific patterns may still cause models to overly rely on these biases, reducing their generalizability to unseen data. These challenges highlight the trade-offs involved in domain-centric approaches, particularly when scaling to more granular or data-scarce tasks.

\end{document}